\documentclass[conference]{IEEEtran_mod}

\usepackage[cmex10]{amsmath}
\usepackage{cite}

\usepackage{latexsym}
\usepackage{graphics}
\usepackage{amssymb}
\usepackage{amsthm}

\usepackage{cite}
\usepackage{array}
\usepackage{graphicx}

\usepackage{url}

\def\tr{\mathrm{tr}}

\def\diag{\mathrm{diag}}

\def\p{d}

\newcounter{MYtempeqncnt}

\newcommand{\fracSumtwo}[2]{\overset{#2}{\underset{#1}{\sum}}}
\newcommand{\vect}[1]{\mathbf{#1}}

\theoremstyle{plain}
\newtheorem{remark}{Remark}

\newtheorem{theorem}{Theorem}
\newtheorem{corollary}{Corollary}
\newtheorem{lemma}{Lemma}
\newtheorem{definition}{Definition}

\begin{document}

\title{Hardware~Impairments~in~{Large-scale}~MISO~Systems: \\ Energy Efficiency, Estimation, and Capacity Limits}

\IEEEoverridecommandlockouts

\author{\IEEEauthorblockN{Emil Bj{\"o}rnson\IEEEauthorrefmark{1}\IEEEauthorrefmark{2}, Jakob Hoydis\IEEEauthorrefmark{4}, Marios Kountouris\IEEEauthorrefmark{3}, and
M{\'e}rouane Debbah\IEEEauthorrefmark{1} \thanks{E. Bj\"ornson is funded by the International Postdoc Grant 2012-228 from The Swedish Research Council. This research has been supported by the ERC Starting Grant 305123 MORE (Advanced Mathematical Tools for Complex Network Engineering). Parts of this work have been performed in the framework of the FP7 project ICT-317669 METIS.}}
\IEEEauthorblockA{\IEEEauthorrefmark{1}Alcatel-Lucent Chair on Flexible Radio, SUPELEC, Gif-sur-Yvette, France (\{emil.bjornson, merouane.debbah\}@supelec.fr)}
\IEEEauthorblockA{\IEEEauthorrefmark{2}ACCESS Linnaeus Centre, Signal Processing Lab, KTH Royal Institute of Technology, Stockholm, Sweden}
\IEEEauthorblockA{\IEEEauthorrefmark{3}Department of Telecommunications, SUPELEC, Gif-sur-Yvette, France (marios.kountouris@supelec.fr)}
\IEEEauthorblockA{\IEEEauthorrefmark{4}Bell Laboratories, Alcatel-Lucent, Stuttgart, Germany (jakob.hoydis@alcatel-lucent.com)}
}

\maketitle

\begin{abstract}
The use of large-scale antenna arrays has the potential to bring substantial improvements in energy efficiency and/or spectral efficiency to future wireless systems, due to the greatly improved spatial beamforming resolution. Recent asymptotic results show that by increasing the number of antennas one can achieve a large array gain and at the same time naturally decorrelate the user channels; thus, the available energy can be focused very accurately at the intended destinations without causing much inter-user interference. Since these results rely on asymptotics, it is important to investigate whether the conventional system models are still reasonable in the asymptotic regimes. This paper analyzes the fundamental limits of large-scale multiple-input single-output (MISO) communication systems using a generalized system model that accounts for transceiver hardware impairments. As opposed to the case of ideal hardware, we show that these practical impairments create finite ceilings on the estimation accuracy and capacity of large-scale MISO systems. Surprisingly, the performance is only limited by the hardware at the single-antenna user terminal, while the impact of impairments at the large-scale array vanishes asymptotically. Furthermore, we show that an arbitrarily high energy efficiency can be achieved by reducing the power while increasing the number of antennas.
\end{abstract}

\IEEEpeerreviewmaketitle

\section{Introduction}

The spectral efficiency in wireless communications is not only limited by the information-theoretic capacity \cite{Telatar1999a}, but also by practical issues such as propagation environment, channel estimation accuracy \cite{Medard2000a}, transceiver hardware impairments \cite{Bjornson2013c}, and signal processing complexity \cite{Rusek2013a}. It is of profound importance to increase the spectral efficiency in future networks, to keep up with the increasing demand for wireless services. However, this is a challenging task and usually comes at the price of having stricter hardware and overhead requirements.

A new network architecture has recently been proposed with the remarkable potential of both increasing the spectral efficiency and taking care of the aforementioned practical issues. It is known as large-scale multiple-input multiple-output (MIMO), or ``massive MIMO'', and is based on having a large number of antennas at the base stations and exploiting channel reciprocity in time-division duplex (TDD) mode \cite{Jose2011b,Hoydis2013a,Ngo2013a,Rusek2013a}. Some key features are: 1) propagation losses are mitigated by a large array gain due to coherent beamforming; 2) impact of channel estimation errors vanishes asymptotically in the large-dimensional space; 3) low-complexity signal processing algorithms are asymptotically optimal; and 4) inter-user interference is easily mitigated by the high beamforming resolution.

The impact of transceiver hardware impairments on large-scale MIMO has received little attention, although large arrays might only be economically feasible if inexpensive hardware can be used at each antenna unit. Cheap hardware components are particularly prone to the impairments (e.g., amplifier non-linearities, I/Q-imbalance, and phase noise) that exist in any physical transceiver implementation \cite{Schenk2008a,Studer2010a,Zetterberg2011a,Bjornson2012b}. Transceiver impairments fundamentally limit the capacity in the high-power regime \cite{Bjornson2013c}, while their impact in the large-antenna regime is less known. To the best of our knowledge, the only previous work in this area is \cite{Pitarokoilis2012a} which derives lower bounds on the achievable uplink sum rate for systems with phase noise.

This paper analyzes the \emph{aggregate} impact of hardware impairments on systems with large antenna arrays, in contrast to the ideal hardware considered in \cite{Jose2011b,Hoydis2013a,Ngo2013a} and the single type of impairments in \cite{Pitarokoilis2012a}.
Section \ref{sec:system-model} reviews the generalized system model with hardware impairments from \cite{Schenk2008a,Studer2010a,Zetterberg2011a,Bjornson2012b}. Section \ref{sec:channel-estimation} derives a new pilot-based channel estimator and shows that the estimation accuracy is limited by the level of impairments. Section \ref{sec:downlink-capacity} derives lower and upper bounds on the downlink capacity and shows that there exists a finite capacity ceiling in large-scale MISO systems with impairments. Despite these discouraging results, Section \ref{sec:energy-efficiency} shows that a very high energy efficiency and resilience towards hardware impairments at the base station can be achieved. For brevity, we consider a single-user system, but extensions to multi-cell scenarios (e.g., with pilot contamination) are outlined throughout the manuscript.

\begin{figure}
\begin{center}
\includegraphics[width=0.82\columnwidth]{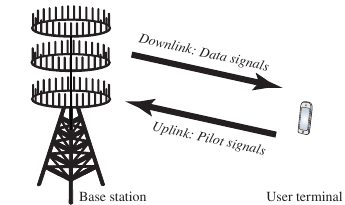}
\end{center} \vskip-3mm
\caption{Example of a large-scale MISO system where the base station has 81 antenna units and can feature $N=162$ antennas using dual polarization.} \label{figure:system-model} \vskip-4mm
\end{figure}

\section{System Model}
\label{sec:system-model}

This paper analyzes the fundamental spectral and energy efficiency limits of a communication system with an $N$-antenna base station (BS) and a single-antenna user terminal (UT). A main characteristic in the analysis will be that the number of antennas $N$ can be very large. To avoid infeasible overhead signaling, we consider a TDD system where channel state information (CSI) is obtained by pilot signaling in the uplink.  The acquired CSI will be utilized for downlink data transmission, by exploiting channel reciprocity; see Fig.~\ref{figure:system-model}.

The flat-fading channel between the base station and the user terminal is denoted $\vect{h} \in \mathbb{C}^{N \times 1}$. This stochastic channel has a static realization for the duration of a data block and independent realizations between blocks. Each realization is circularly-symmetric complex Gaussian: $\vect{h} \sim \mathcal{CN}(\vect{0},\vect{R})$. This is known as Rayleigh fading and the positive semi-definite covariance matrix is $\mathbb{E}\{ \vect{h} \vect{h}^H\} = \vect{R} \in \mathbb{C}^{N \times N}$, where  $\mathbb{E}\{\cdot\}$ denotes expectation. We assume that the spectral norm of $\vect{R}$ is uniformly bounded, irrespective of the number of antennas $N$. This is a common technical assumption that enables asymptotic analysis (cf.~\cite{Hoydis2013a}), but which is also a necessary physical property that originates from the law of energy conservation.

Physical transceivers suffer from hardware impairments that 1) create a mismatch between the intended data signal and what is actually generated and emitted; and 2) distort the received signal in the reception processing. This is modeled herein by a generalized channel model from \cite{Schenk2008a,Studer2010a,Zetterberg2011a,Bjornson2012b} with additive distortion noises at the transmitter and the receiver.

\subsection{Generalized Downlink Channel Model}

The downlink channel is used for data transmission; see Fig.~\ref{figure:system-model}. The received downlink signal $y \in \mathbb{C}$ in flat-fading MISO systems is conventionally modeled as
\begin{equation} \label{eq:downlink_channel-classic}
y = \vect{h}^T \vect{w} s + n
\end{equation}
where $s$ is the random zero-mean data signal with 
 power $\mathbb{E}\{ |s|^2 \} = p^{\mathrm{BS}}$ and $\vect{w} \in \mathbb{C}^{N \times 1}$ is the unit-norm beamforming vector.
The receiver noise is modeled as $n \sim \mathcal{CN}(0,\sigma_{\mathrm{UT}}^2)$ and might include interference from neighboring systems.

To model physical channels more accurately, this paper considers a generalized model \cite{Schenk2008a,Studer2010a,Zetterberg2011a,Bjornson2012b} where the received signal is
\begin{equation} \label{eq:downlink_channel}
y = \vect{h}^T (\vect{w} s + \boldsymbol{\eta}_{t}^{\mathrm{BS}}) + n + \eta_{r}^{\mathrm{UT}}.
\end{equation}
The difference from the conventional model \eqref{eq:downlink_channel-classic} is the additional distortion noise terms $\boldsymbol{\eta}_{t}^{\mathrm{BS}} \sim \mathcal{CN}(\vect{0},\vect{\Upsilon}_t^{\mathrm{BS}})$ and $\eta_{r}^{\mathrm{UT}} \sim \mathcal{CN}(0,\upsilon_r^{\mathrm{UT}})$. These are independent and describe the impact of impairments in the transmitter and receiver hardware, respectively \cite{Schenk2008a}. Theoretical investigations and several measurements have suggested that
\begin{equation} \label{eq:distortion-statistics-DL}
\begin{split}
\vect{\Upsilon}_t^{\mathrm{BS}} &= \kappa_t^{\mathrm{BS}} p^{\mathrm{BS}} \, \diag(|w_1|^2,\ldots,|w_{N}|^2) \\
\upsilon_r^{\mathrm{UT}} &= \kappa_r^{\mathrm{UT}} p^{\mathrm{BS}} \, |\vect{h}^T \vect{w}|^2
\end{split}
\end{equation}
where $w_i$ is the $i$th element of $\vect{w}$ and $\diag(\cdot,\ldots,\cdot)$ denotes a diagonal matrix. The Gaussianity is explained by the aggregate effect of many impairments and the use of compensation algorithms that remove other types of distortions \cite{Studer2010a,Zetterberg2011a}.

\begin{remark}
Distortion noise is an alteration of the data signal, while the classical receiver noise models random fluctuations in the electronic circuits at the receiver. A main difference is thus that the average distortion noise power is proportional to the signal power $p^{\mathrm{BS}}$ and the current channel gain $\|\vect{h}\|^2$.\footnote{The total transmit power is $p^{\mathrm{BS}} (1+\kappa_t^{\mathrm{BS}})$ under hardware impairments and not the usual $p^{\mathrm{BS}}$, since distortions also contribute a small amount of power. For simplicity, we will still refer to the signal power $p^{\mathrm{BS}}$ as the transmit power since the total power is fully characterized by and is very close to $p^{\mathrm{BS}}$.} The proportionality constants $\kappa_t^{\mathrm{BS}},\kappa_r^{\mathrm{UT}} \geq 0$ characterize the \emph{level of impairments} and are typically in the range $[0,0.03]$.\footnote{The square root of each $\kappa$-parameter equals the error vector magnitude (EVM), which is a common quality measure of transceivers.} Small values represent accurate and expensive transceiver hardware.
\end{remark}

\subsection{Generalized Uplink Channel Model}

The reciprocal uplink channel will be used for pilot-based channel estimation; see Fig.~\ref{figure:system-model}.
Similar to \eqref{eq:downlink_channel}, we consider a generalized model with the received signal $\vect{z} \in \mathbb{C}^{N}$ being
\begin{equation} \label{eq:uplink_channel}
\vect{z} = \vect{h} (\p + \eta_{t}^{\mathrm{UT}}) + \boldsymbol{\nu} + \boldsymbol{\eta}_{r}^{\mathrm{BS}}
\end{equation}
where $\p \in \mathbb{C}$ is a scalar deterministic pilot signal with power $p^{\mathrm{UT}} = |\p|^2$. The independent additive noise $\boldsymbol{\nu} \sim \mathcal{CN}(\vect{0},\vect{S})$ has a positive definite covariance matrix $\vect{S} \in \mathbb{C}^{N \times N}$.
This term contains both receiver noise and potential interference from neighboring systems (e.g., pilot contamination \cite{Jose2011b,Hoydis2013a,Ngo2013a,Rusek2013a}). We assume that $\vect{S}$ has a uniformly bounded spectral norm as $N \rightarrow \infty$, for the same physical reasons as for $\vect{R}$.

The transceiver impairments in the hardware used for uplink transmission are modeled by the independent distortion noises $\eta_{t}^{\mathrm{UT}} \sim \mathcal{CN}(0,\upsilon_t^{\mathrm{UT}})$ and $\boldsymbol{\eta}_{r}^{\mathrm{BS}} \sim \mathcal{CN}(\vect{0},\vect{\Upsilon}_r^{\mathrm{BS}})$ at the transmitter and receiver, respectively. Similar to \eqref{eq:distortion-statistics-DL}, their covariance matrices are model as
\begin{equation} \label{eq:distortion-statistics-UL}
\begin{split}
\vect{\Upsilon}_r^{\mathrm{BS}} &= \kappa_r^{\mathrm{BS}} p^{\mathrm{UT}} \, \diag(|h_1|^2,\ldots,|h_{N}|^2) \\
\upsilon_t^{\mathrm{UT}} &= \kappa_t^{\mathrm{UT}} p^{\mathrm{UT}}. \\
\end{split}
\end{equation}
Note that we might have $\kappa_t^{\mathrm{BS}} \neq \kappa_r^{\mathrm{BS}}$ at the base station and $\kappa_r^{\mathrm{UT}} \neq\kappa_t^{\mathrm{UT}}$ at the user terminal since different pieces of hardware are generally used for transmission and reception.

\section{Uplink Channel Estimation}
\label{sec:channel-estimation}

This section considers estimation of the current channel realization $\vect{h}$ by comparing the received uplink signal $\vect{z}$ in \eqref{eq:uplink_channel} with the predefined pilot signal $\p$.
The classic results on pilot-based channel estimation consider Rayleigh fading channels that are observed in independent complex Gaussian noise with known statistics \cite{Kay1993a,Bjornson2010a}. However, this is not the case herein because the distortion noises $\eta_{t}^{\mathrm{UT}}$ and $\boldsymbol{\eta}_{r}^{\mathrm{BS}}$ depend on the unknown random channel $\vect{h}$. The dependence is either through the multiplication $\vect{h} \eta_{t}^{\mathrm{UT}}$ or the conditional variance of $\boldsymbol{\eta}_{r}^{\mathrm{BS}}$ in  \eqref{eq:distortion-statistics-UL}, which is essentially the same relation. Although the distortion noises are Gaussian when conditioning on a channel realization, the effective distortion is the product of Gaussian variables and, thus, has a \emph{complex double Gaussian} distribution \cite{Donoughue2012a}.
Consequently, an optimal channel estimator cannot be deduced from the standard results provided in \cite{Kay1993a,Bjornson2010a}.

We will now derive the \emph{linear} minimum mean square error (LMMSE) estimator. Note that $()^*$ denotes conjugation.

\begin{theorem} \label{theorem:LMMSE-estimator}
The LMMSE estimator of $\vect{h}$ from the observation of $\vect{z}$ in \eqref{eq:uplink_channel} is
\begin{equation} \label{eq:LMMSE-estimator}
\begin{split}
\widehat{\vect{h}} = \p^* \vect{R} \Big( p^{\mathrm{UT}} (1+\kappa_t^{\mathrm{UT}}) \vect{R} + p^{\mathrm{UT}} \kappa_r^{\mathrm{BS}}  \vect{R}_{\diag} + \vect{S} \Big)^{-1} \vect{z}
\end{split}
\end{equation}
where $\vect{R}_{\diag} = \diag(r_{11},\ldots,r_{NN})$ consists of the diagonal elements of $\vect{R}$. The total $\textrm{MSE} = \mathbb{E}\{ \|\widehat{\vect{h}}  - \vect{h}\|^2 \} $ is $\tr( \vect{C} )$, where $\vect{C} = \mathbb{E}\{ (\widehat{\vect{h}}  - \vect{h})(\widehat{\vect{h}}  - \vect{h})^H \}$ is the error covariance matrix:
\begin{equation} \label{eq:LMMSE-error-covariance}
\vect{C} =  \vect{R} - p^{\mathrm{UT}} \vect{R} \left( p^{\mathrm{UT}} (1+\kappa_t^{\mathrm{UT}}) \vect{R} + p^{\mathrm{UT}} \kappa_r^{\mathrm{BS}}  \vect{R}_{\diag} + \vect{S} \right)^{-1} \vect{R}.
\end{equation}
\end{theorem}
\begin{IEEEproof}
The LMMSE estimator has the form $\widehat{\vect{h}} = \vect{A} \vect{z}$ where $\vect{A}$ minimizes  $\textrm{MSE} =  \mathbb{E}\{ \|\widehat{\vect{h}}  - \vect{h}\|^2 \} $. An expression is obtained by taking expectations over $\eta_t^{\mathrm{UT}},\boldsymbol{\eta}_r^{\mathrm{BS}}$ for an arbitrary fixed $\vect{h}$ and then taking the expectation over $\vect{h}$. The LMMSE estimator in \eqref{eq:LMMSE-estimator} follows from straightforward differentiation and $\vect{C}$ and the MSE then follow from the definitions.
\end{IEEEproof}

We remark that there might exist non-linear estimators that outperform the LMMSE estimator in Theorem \ref{theorem:LMMSE-estimator}. This stands in contrast to conventional channel estimation with independent Gaussian noise, where the LMMSE estimator is also the MMSE estimator \cite[Remark 1]{Bjornson2010a}. However, the difference in MSE performance should be small, since the dependent distortion noises are relatively weak.

\begin{corollary} \label{cor:asymptotic-estimation-performance}
Consider the special case of $\vect{R} = \lambda \vect{I}$ and $\vect{S} = \sigma_{\mathrm{BS}}^2 \vect{I}$. The error covariance matrix in \eqref{eq:LMMSE-error-covariance} becomes
\begin{equation} \label{eq:LMMSE-error-covariance-iid}
\vect{C} =  \lambda \left( 1  - \frac{ p^{\mathrm{UT}} \lambda }{  p^{\mathrm{UT}} \lambda (1 + \kappa_t^{\mathrm{UT}}  +\kappa_r^{\mathrm{BS}}  ) + \sigma_{\mathrm{BS}}^2 }  \right) \vect{I}.
\end{equation}
In the high uplink power\footnote{The level of impairments in the transmitter hardware increases with the transmit power \cite{Bjornson2012b}. This means that $\kappa_t^{\mathrm{BS}},\kappa_t^{\mathrm{UT}}$ are almost constant within the dynamic range of the power amplifier, but increase rapidly when moving outside this range. This is not taken into account, since the high-power regime is not our main focus. Consequently, the high-power limits derived herein are optimistic and might not be achievable in practice. The results when $N \rightarrow \infty$ are however accurate since this means that less power is allocated per antenna.} regime, we have
\begin{equation} \label{eq:LMMSE-error-covariance-iid-asympt}
\lim_{p^{\mathrm{UT}} \rightarrow \infty } \vect{C} = \lambda \left( 1 - \frac{1}{1 + \kappa_t^{\mathrm{UT}}  +\kappa_r^{\mathrm{BS}}} \right) \vect{I}.
\end{equation}
\end{corollary}

This corollary brings important insights on the average estimation error per element in $\vect{h}$. The error variance is given by the factor in front of the identity matrix in \eqref{eq:LMMSE-error-covariance-iid}. It is independent of the number of antennas $N$, thus letting $N$ grow large will neither increase nor decrease the estimation error \emph{per element}.\footnote{The MSE per element is finite, $\frac{1}{N} \tr(\vect{C}) < \infty$, but the sum MSE behaves as $\tr( \vect{C} ) \rightarrow \infty$ when $N \rightarrow \infty$ since the number of elements increases.} The estimation error is clearly a decreasing function of the pilot power $p^{\mathrm{UT}}$, however the error variance will \emph{not} converge to zero as the pilot power is increased. As seen in \eqref{eq:LMMSE-error-covariance-iid-asympt}, there is a strictly positive error floor of $\lambda ( 1 -\frac{1}{1 + \kappa_t^{\mathrm{UT}}  +\kappa_r^{\mathrm{BS}}})$ due to the transceiver hardware impairments. Thus, perfect estimation accuracy cannot be achieved in practice, not even asymptotically. The error floor is characterized by the sum of the levels of impairments $\kappa_t^{\mathrm{UT}},\kappa_r^{\mathrm{BS}}$ in the transmitter and receiver hardware, respectively. In terms of estimation accuracy, it is therefore equally important to have high-quality hardware at both the base station and the user terminal.

There will be an error floor also for non-diagonal $\vect{R}$ and $\vect{S}$; the general high-power limit is easily computed from \eqref{eq:LMMSE-error-covariance}.

The channel can be decomposed as $\vect{h} = \widehat{\vect{h}} + \boldsymbol{\epsilon}$, where $\widehat{\vect{h}}$ is the LMMSE estimate in Theorem \ref{theorem:LMMSE-estimator} and $\boldsymbol{\epsilon} \in \mathbb{C}^{N \times 1}$ denotes the unknown estimation error. Contrary to conventional estimation with independent noise (cf.~\cite[Chapter 15.8]{Kay1993a}), $\widehat{\vect{h}}$ and $\boldsymbol{\epsilon}$ are neither independent nor jointly complex Gaussian, but only uncorrelated and zero-mean. The covariance matrices are
$\mathbb{E}\{ \widehat{\vect{h}} \widehat{\vect{h}}^H\} = \vect{R} - \vect{C}$ and $\mathbb{E}\{ \boldsymbol{\epsilon} \boldsymbol{\epsilon}^H\} = \vect{C}$ with $\vect{C}$ given in \eqref{eq:LMMSE-error-covariance}.

\subsection{Numerical Illustration}

\begin{figure}
\begin{center}
\includegraphics[width=\columnwidth]{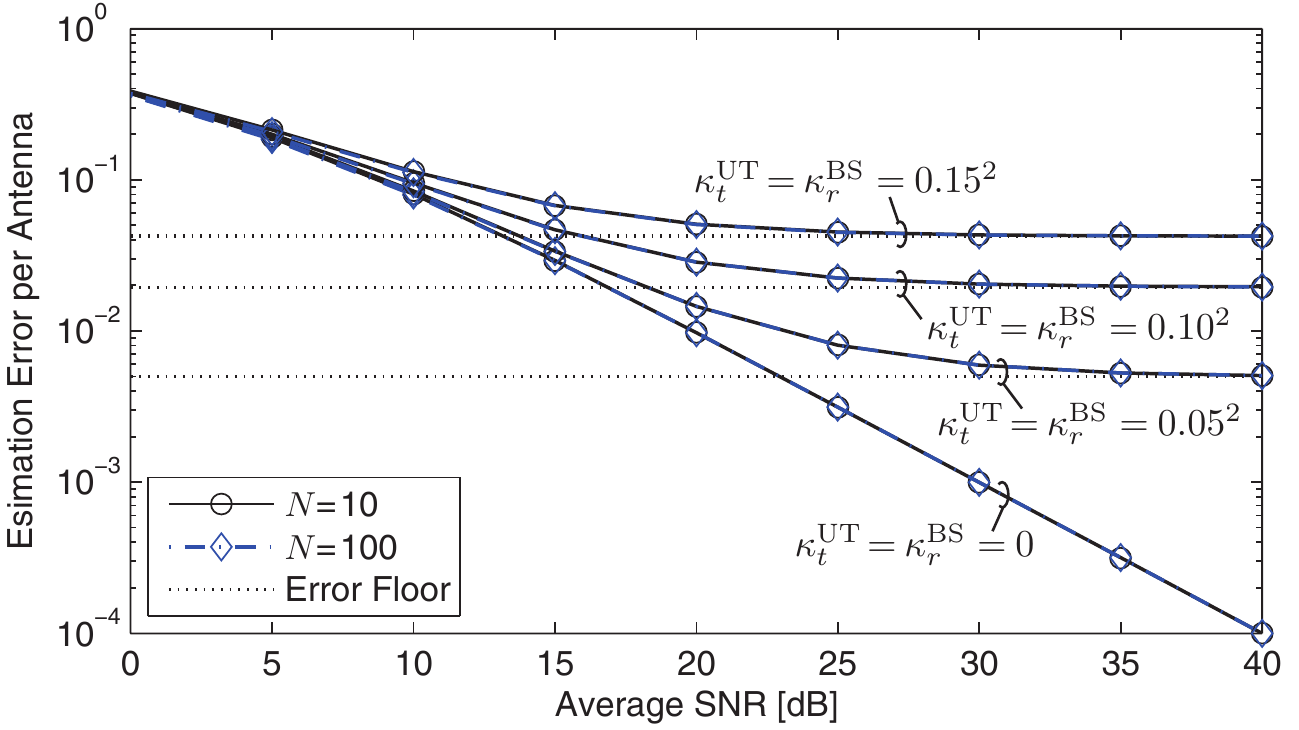} \vskip-2mm
\caption{Estimation error per antenna element for the LMMSE estimator in Theorem \ref{theorem:LMMSE-estimator}. Transceiver hardware impairments create a non-zero error floor.}\label{figure_estimation_error}
\end{center} \vskip-4mm
\end{figure}

The estimation accuracy of the LMMSE estimator under hardware impairments is illustrated in Fig.~\ref{figure_estimation_error} with $N=10$ and $N=100$ antennas. The channel covariance matrices $\vect{R}$  are generated using the exponential model with correlation coefficient 0.7 \cite{Loyka2001a}, while the noise covariance is $\vect{S} = \vect{I}$. Fig.~\ref{figure_estimation_error} shows the (normalized) estimation error per channel element as a function of the average SNR, defined as $p^{\mathrm{UT}} \frac{\tr (\vect{R})}{\tr (\vect{S})}$.

Four hardware setups with different levels of impairments are considered: $\kappa_t^{\mathrm{UT}} =\kappa_r^{\mathrm{BS}} \in \{ 0, 0.05^2, 0.10^2, 0.15^2 \}$. Fig.~\ref{figure_estimation_error} confirms that there are non-zero error floors at high SNRs, as proved by Corollary \ref{cor:asymptotic-estimation-performance}. The error floor increases when increasing the level of impairments. The estimation error is very close to the floor when the uplink SNR reaches 20-30 dB, thus further increase in SNR only brings minor improvement.
Moreover, the curves for $N=10$ and $N=100$ are virtually the same, showing that the error \emph{per} antenna is almost independent of $N$ even under high spatial channel correlation.

\section{Downlink Data Transmission}
\label{sec:downlink-capacity}

This section analyzes the capacity of the downlink channel in \eqref{eq:downlink_channel}, where a linear beamforming vector $\vect{w}$ is applied to transform the MISO channel $\vect{h}$ into an effective single-input single-output (SISO) channel $\vect{h}^T \vect{w} \in \mathbb{C}$. The beamforming vector is a function of the estimate $\hat{\vect{h}}$ obtained at the base station and the statistical properties of other channel parameters.

The receiver has a statistical characterization of the effective channel $\vect{h}^T \vect{w}$, obtained from the long-term statistics and possibly some additional downlink pilot signaling. We leave the extent of CSI at the user terminal undefined since the bounds derived in this section hold for anything between no downlink pilot signaling and perfect CSI estimation.

Under these conditions, the capacity of the generalized memoryless fading downlink channel in \eqref{eq:downlink_channel} is
\begin{equation} \label{eq:downlink_capacity}
{\tt C} = \mathbb{E} \left\{ \max_{ \vect{w}(\hat{\vect{h}}) \,: \, \| \vect{w}\|=1 }\, \, \max_{f(s) \,: \, \mathbb{E}\{|s|^2\} \leq p^{\mathrm{BS}}}  \mathcal{I}(s; y)   \right\}
\end{equation}
where $\mathcal{I}(s; y )$ denotes the mutual information between the received signal $y$ and data signal $s$. The beamforming vector $\vect{w}(\hat{\vect{h}})$ can be any deterministic function of $\hat{\vect{h}}$, while $f(s)$ denotes the probability distribution of the data signal. The expression in \eqref{eq:downlink_capacity} combines the well-known capacity formulas from \cite{Caire1999a} with a maximization over $\vect{w}(\hat{\vect{h}})$. The expectation is taken over different data blocks with static values on $\vect{h},\hat{\vect{h}}$.

We will characterize the capacity by deriving lower and upper bounds on \eqref{eq:downlink_capacity} and study their behavior in the limit of infinitely many transmit antennas $N$. This brings insights on how hardware impairments affect large-scale MISO systems.

\subsection{Upper Bound on Channel Capacity}

An upper bound on \eqref{eq:downlink_capacity} can be achieved by adding side information at the transmitter and receiver. We assume that the channel realization $\vect{h}$ is known perfectly at the transmitter and receiver in each fading block.  Since the receiver and distortion noise in \eqref{eq:downlink_channel} are complex Gaussian and independent of the useful signal under perfect CSI, we deduce that Gaussian signaling, $s \sim \mathcal{CN}(0,p^{\mathrm{BS}})$, is optimal \cite{Telatar1999a,Bjornson2013c} and obtain
\begin{align} \label{eq:downlink_capacity_upper1}
&{\tt C} \leq  \mathbb{E} \left\{ \max_{ \vect{w}(\vect{h}) \, : \,  \| \vect{w}\|=1} \log_2( 1+ {\tt SINR}(\vect{w}) ) \right\} \quad \textrm{where}\\
\label{eq:downlink_SINR_upper1}
&{\tt SINR}(\vect{w}) = \frac{ |\vect{h}^T \vect{w}|^2}{ \kappa_t^{\mathrm{BS}}  \fracSumtwo{i=1}{N} |h_i w_i|^2 + \kappa_r^{\mathrm{UT}} |\vect{h}^T \vect{w}|^2   + \frac{\sigma_{\mathrm{UT}}^2}{p^{\mathrm{BS}}} }.
\end{align}

To find an upper bound in closed form, we first need to find the optimal beamforming vector in the current upper bound.

\begin{lemma} \label{lemma:maximizing_beamforming}
The SINR in \eqref{eq:downlink_SINR_upper1} is maximized as
\begin{equation}
 \max_{ \vect{w} \, : \,  \| \vect{w}\|=1} {\tt SINR}(\vect{w}) = \vect{h}^T \Big( \kappa_t^{\mathrm{BS}} \vect{D}_{\vect{h}} + \kappa_r^{\mathrm{UT}} \vect{h}^* \vect{h}^T + \frac{\sigma_{\mathrm{UT}}^2}{p^{\mathrm{BS}}} \vect{I} \Big)^{-1} \vect{h}^*
\end{equation}
for $\vect{D}_{\vect{h}} = \diag(|h_1|^2,\ldots,|h_N|^2)$. The optimum is attained by
\begin{equation} \label{eq:optimal_beamforming_perfectCSI}
\begin{split}
\vect{w} =
\frac{ ( \kappa_t^{\mathrm{BS}} \vect{D}_{\vect{h}} + \frac{\sigma_{\mathrm{UT}}^2}{p^{\mathrm{BS}}} \vect{I} )^{-1} \vect{h}^* }{\big\| \big( \kappa_t^{\mathrm{BS}} \vect{D}_{\vect{h}} + \frac{\sigma_{\mathrm{UT}}^2}{p^{\mathrm{BS}}} \vect{I} \big)^{-1} \vect{h}^* \big\|}.
\end{split}
\end{equation}
\end{lemma}
\begin{IEEEproof}
The SINR can be rewritten as a generalized Rayleigh quotient and is therefore maximized by \eqref{eq:optimal_beamforming_perfectCSI}.
\end{IEEEproof}

Note that the beamforming vector $\vect{w}$ in \eqref{eq:optimal_beamforming_perfectCSI} only depends on impairments at the base station.
Lemma \ref{lemma:maximizing_beamforming} enables us to derive a closed-form upper bound on the channel capacity.

\begin{theorem} \label{theorem:upperbound-capacity}
The capacity in \eqref{eq:downlink_capacity} is upper bounded as
\begin{equation} \label{eq:downlink_capacity_upper2}
{\tt C} \leq {\tt C}_{\mathrm{upper}} = \log_2\left( 1 + \frac{G }{1 +\kappa_r^{\mathrm{UT}} G} \right)
\end{equation}
where $r_{11},\ldots,r_{NN}$ are the diagonal elements of $\vect{R}$,
\begin{equation} \label{eq:expression_G}
G = \sum_{i=1}^{N} \frac{1}{\kappa_t^{\mathrm{BS}}} \! \left( 1 - \frac{\sigma_{\mathrm{UT}}^2}{p^{\mathrm{BS}} \kappa_t^{\mathrm{BS}} r_{ii}} E_1 \bigg( \frac{\sigma_{\mathrm{UT}}^2}{p^{\mathrm{BS}} \kappa_t^{\mathrm{BS}} r_{ii}} \bigg) e^{\frac{\sigma_{\mathrm{UT}}^2}{p^{\mathrm{BS}} \kappa_t^{\mathrm{BS}} r_{ii}}} \right),
\end{equation}
and $E_1(x) = \int_{1}^{\infty} \frac{e^{-tx} }{t} dt$ denotes the exponential integral.
\end{theorem}
\begin{IEEEproof}
Using Lemma \ref{lemma:maximizing_beamforming} and the Woodbury identity,
the upper bound in \eqref{eq:downlink_capacity_upper1} can be expressed as $\mathbb{E}\{ m(\psi) \} = \mathbb{E}\{ \log_2(1+\frac{\psi}{1+\kappa_r^{\mathrm{UT}} \psi} ) \}$, where $\psi= \vect{h}^T ( \kappa_t^{\mathrm{BS}} \vect{D}_{\vect{h}} + \frac{\sigma_{\mathrm{UT}}^2}{p^{\mathrm{BS}}} \vect{I} )^{-1} \vect{h}^*$. Since $m(\cdot)$ is a concave function, we apply Jensen's inequality to obtain a new upper bound $\mathbb{E} \left\{ m(\psi) \right\} \leq m(\mathbb{E} \left\{ \psi \right\} )$. Finally, \eqref{eq:downlink_capacity_upper2} and \eqref{eq:expression_G} are obtained by evaluating $\mathbb{E} \left\{ \psi \right\}$.
\end{IEEEproof}

This closed-form upper bound provides important insights on the achievable performance under hardware impairments.

\begin{figure*}[!t]
\normalsize
\setcounter{MYtempeqncnt}{\value{equation}}

\setcounter{equation}{20}
\begin{equation} \label{eq:capacity-lower-equivalent}
{\tt C} \!\geq\! {\tt C}_{\mathrm{lower}}\! =\!
 \log_2 \! \left(  \! 1 \!+\! \frac{  \bigg|  \mathbb{E}\bigg\{ \!
\frac{(1+\p^{-1} \eta_{t}^{\mathrm{UT}}) \sqrt{\tr(\vect{R}-\vect{C})} }{\sqrt{ \tr\big(\vect{A} ( |\p + \eta_{t}^{\mathrm{UT}}|^2 \vect{R} + \boldsymbol{\Psi}) \vect{A}^H \big) }}
\! \bigg\} \bigg|^2 + \mathcal{O} \left(\frac{1}{\sqrt{N}} \right)
 }{ (1 \!+ \! \kappa_r^{\mathrm{UT}})
   \mathbb{E}\bigg\{ \!
\frac{ |1+\p^{-1} \eta_{t}^{\mathrm{UT}} |^2 \tr(\vect{R}-\vect{C}) }{ \tr \big( \vect{A} ( |\p + \eta_{t}^{\mathrm{UT}}|^2 \vect{R} + \boldsymbol{\Psi}) \vect{A}^H \big) } \!
\bigg\} \!-\!  \bigg|  \mathbb{E}\bigg\{ \!
\frac{(1+\p^{-1} \eta_{t}^{\mathrm{UT}}) \sqrt{\tr(\vect{R}-\vect{C})} }{\sqrt{ \tr\big(\vect{A} ( |\p + \eta_{t}^{\mathrm{UT}}|^2 \vect{R} + \boldsymbol{\Psi}) \vect{A}^H \big) }}
\! \bigg\} \bigg|^2  \! + \frac{1}{N}\frac{\sigma_{\mathrm{UT}}^2}{p^{\mathrm{UT}} p^{\mathrm{BS}}} + \mathcal{O} \left(\frac{1}{\sqrt{N}} \right) } \! \right)
\end{equation}
\setcounter{equation}{\value{MYtempeqncnt}} \vskip-4mm
\hrulefill
\vskip-3mm
\end{figure*}

\begin{corollary} \label{cor:upper_bound}
The upper capacity bound in  \eqref{eq:downlink_capacity_upper2} has the following asymptotic properties:
\begin{align}
\lim_{p^{\mathrm{BS}} \rightarrow \infty} {\tt C}_{\mathrm{upper}} &= \log_2\left( 1 + \frac{N}{\kappa_t^{\mathrm{BS}} +\kappa_r^{\mathrm{UT}} N} \right) \label{eq:upper-asymptotics-q}\\
\lim_{N \rightarrow \infty } {\tt C}_{\mathrm{upper}} &= \log_2\left( 1 + \frac{1}{\kappa_r^{\mathrm{UT}}} \right) \label{eq:upper-asymptotics-N}.
\end{align}
\end{corollary}
\begin{IEEEproof}
Note that $r_1>0,\ldots,r_N>0$ since $\vect{R}$ is positive definite.
This implies $G \rightarrow \sum_{i=1}^N \frac{1}{\kappa_t^{\mathrm{BS}}} = \frac{N}{\kappa_t^{\mathrm{BS}}}$ as $p^{\mathrm{BS}} \rightarrow \infty$ for fixed $N$, giving \eqref{eq:upper-asymptotics-q}. It also implies $\frac{1}{N} G > 0$ as $N \rightarrow \infty$, thus $\frac{G }{1 +\kappa_r^{\mathrm{UT}} G}  - \frac{G }{\kappa_r^{\mathrm{UT}} G} \rightarrow 0$ as $N \rightarrow \infty$ giving \eqref{eq:upper-asymptotics-N}.
\end{IEEEproof}

This corollary shows that the capacity has finite ceilings both when the downlink transmit power $p^{\mathrm{BS}}$ grows large and when the number of antennas $N$ grows large. The ceilings depend on the impairment parameters $\kappa_t^{\mathrm{BS}},\kappa_r^{\mathrm{UT}}$ and the UT impairments are clearly $N$ times more influential. Note that even very small hardware impairments will ultimately limit the capacity. In other words, the ever-increasing capacity observed in the high-SNR and large-$N$ regimes with perfect transceiver hardware (cf.~\cite{Jose2011b,Hoydis2013a,Ngo2013a,Rusek2013a}) is not easily achieved in practice.

Interestingly, \eqref{eq:upper-asymptotics-N} indicates that only the quality of the terminal's receiver hardware limits the capacity as $N \rightarrow \infty$. This means that the detrimental effect of hardware impairments in the base station disappears when the number of base station antennas grow large. This is, simply speaking, since the distortion noise is spread isotropically in space while the increased spatial resolution of the array enables very exact transmit beamforming. This is a very promising result since large arrays are more prone to impairments, due to economic and implementation limitations. To verify this observation, we also need a lower bound on the channel capacity.

\subsection{Lower Bound on Channel Capacity}

We obtain a lower capacity bound by making the potentially limiting assumptions of Gaussian codebook, no downlink pilot signaling, and Gaussian CSI uncertainty at the terminal. The next theorem is obtained by an approach from \cite{Medard2000a}, which has been applied to massive MIMO in \cite{Jose2011b,Hoydis2013a,Ngo2013a} (among others).

\begin{theorem} \label{theorem:lower-bound-on-capacity}
The capacity in \eqref{eq:downlink_capacity} is lower bounded as
\begin{equation} \label{eq:lower-bound-capacity}
{\tt C} \geq {\tt C}_{\mathrm{lower}} = \log_2 \left( 1 + \widetilde{{\tt SINR}} \right)
\end{equation}
where $\vect{v} = [v_1 \, \ldots \, v_{K_r}]^T$ with $\| \vect{v}\|=1$ is a function of $\hat{\vect{h}}$ and
\begin{equation} \label{eq:SINR-approximation-lower-bound}
\begin{split}
&\widetilde{{\tt SINR}} = \\[-2mm]
&\frac{ \left| \mathbb{E}\{ \vect{h}^T \vect{v} \}  \right|^2 }{  (\! 1\!+\! \kappa_r^{\mathrm{UT}} \!) \mathbb{E} \left\{ | \vect{h}^T \vect{v} |^2 \! \right\} \!-\! \left| \mathbb{E}\{ \vect{h}^T \vect{v} \}  \! \right|^2 \!+\! \kappa_t^{\mathrm{BS}} \! \fracSumtwo{i=1}{N} \mathbb{E}\{ |h_i|^2 |v_i|^2\} \!+\! \frac{\sigma_{\mathrm{UT}}^2}{p^{\mathrm{BS}}} }.
\end{split}
\end{equation}
\end{theorem} \vskip-3mm
\begin{IEEEproof}
Any heuristic beamforming $\vect{w}(\hat{\vect{h}}) = \vect{v}$ gives a lower bound. This transforms \eqref{eq:downlink_capacity} into a SISO channel and we use the approach in \cite[Section III]{Medard2000a} to find a lower bound on capacity with only statistics of $\vect{h}^T \vect{v}$ at the receiver.
\end{IEEEproof}

The lower bound in \eqref{eq:lower-bound-capacity} can be computed numerically for any channel distribution and any way of selecting the beamforming vector based on the channel estimate. To bring some insight on the behavior when the number of antennas, $N$, grows large, we have the following result for the case of (approximate) maximum ratio transmission (MRT).

\begin{corollary} \label{cor:lower_bound}
Suppose $\vect{v}= \frac{\hat{\vect{h}}^*}{\|\hat{\vect{h}}\|}$, then the lower bound in \eqref{eq:lower-bound-capacity} can be expressed as in \eqref{eq:capacity-lower-equivalent} at the top of the next page. The big-$\mathcal{O}$ terms $\mathcal{O} \left(\frac{1}{\sqrt{N}} \right)$  vanish as $\frac{1}{\sqrt{N}}$ when $N \rightarrow \infty$, while the other terms (except the noise term) remain strictly positive.
\end{corollary}
\begin{IEEEproof}
The equivalence between \eqref{eq:lower-bound-capacity} and \eqref{eq:capacity-lower-equivalent} is proved by straightforward but lengthy use of H\"older's inequality and the trace lemma in
\cite[Lemma B.26]{Bai2009a}.
\end{IEEEproof}

Combining the upper bound in Corollary \ref{cor:upper_bound} with the lower bound in Corollary \ref{cor:lower_bound}, we have a clear characterization of the capacity behavior when $N \rightarrow \infty$. Contrary to the upper bound, the hardware impairments at the base station are present in \eqref{eq:capacity-lower-equivalent} through $\vect{A}$ and $\boldsymbol{\Psi}$. However, these variables vanish if we set $\eta_{t}^{\mathrm{UT}}=0$, in which case the lower bound actually equals the upper bound.
Consequently, the capacity limit is mainly determined by the level of impairments at the user terminal, both in the uplink estimation ($\kappa_{t}^{\mathrm{UT}}$) and the downlink transmission ($\kappa_r^{\mathrm{UT}}$)---although the former connection was not visible in the upper bound since it assumed perfect CSI. 

\begin{remark}
It is implicitly assumed in Corollaries \ref{cor:upper_bound} and \ref{cor:lower_bound} that $\frac{1}{N} \frac{\sigma_{\mathrm{UT}}^2}{p^{\mathrm{UT}} p^{\mathrm{BS}}} \rightarrow 0$ as $N \rightarrow \infty$, meaning that the noise variance $\sigma_{\mathrm{UT}}^2$ cannot scale as fast as $\mathcal{O}(N)$. This assumption generally holds, but not under pilot contamination of the type considered in \cite{Jose2011b,Hoydis2013a,Ngo2013a,Rusek2013a}. In those cases there will be additional terms in the denominator of the SINRs that further limits the capacity. 
\end{remark}

\subsection{Numerical Illustration}

This section illustrates our lower and upper bounds on the capacity.
The average SNR is defined as $p^{\mathrm{UT}} \frac{\tr (\vect{R})}{\tr (\vect{S})}$  and $p^{\mathrm{BS}} \frac{\tr (\vect{R})}{\tr (\vect{S})}$ for pilot and data transmission, respectively, and is fixed at 20 dB, while we vary the number of antennas $N$ and the levels of impairments. For simplicity, $\kappa^{\mathrm{BS}} \triangleq \kappa_t^{\mathrm{BS}} =\kappa_r^{\mathrm{BS}} $ at the base station and $\kappa^{\mathrm{UT}} \triangleq \kappa_t^{\mathrm{UT}} =\kappa_r^{\mathrm{UT}} $ at the user terminal.

\begin{figure} \vskip-1mm
\begin{center}
\includegraphics[width=\columnwidth]{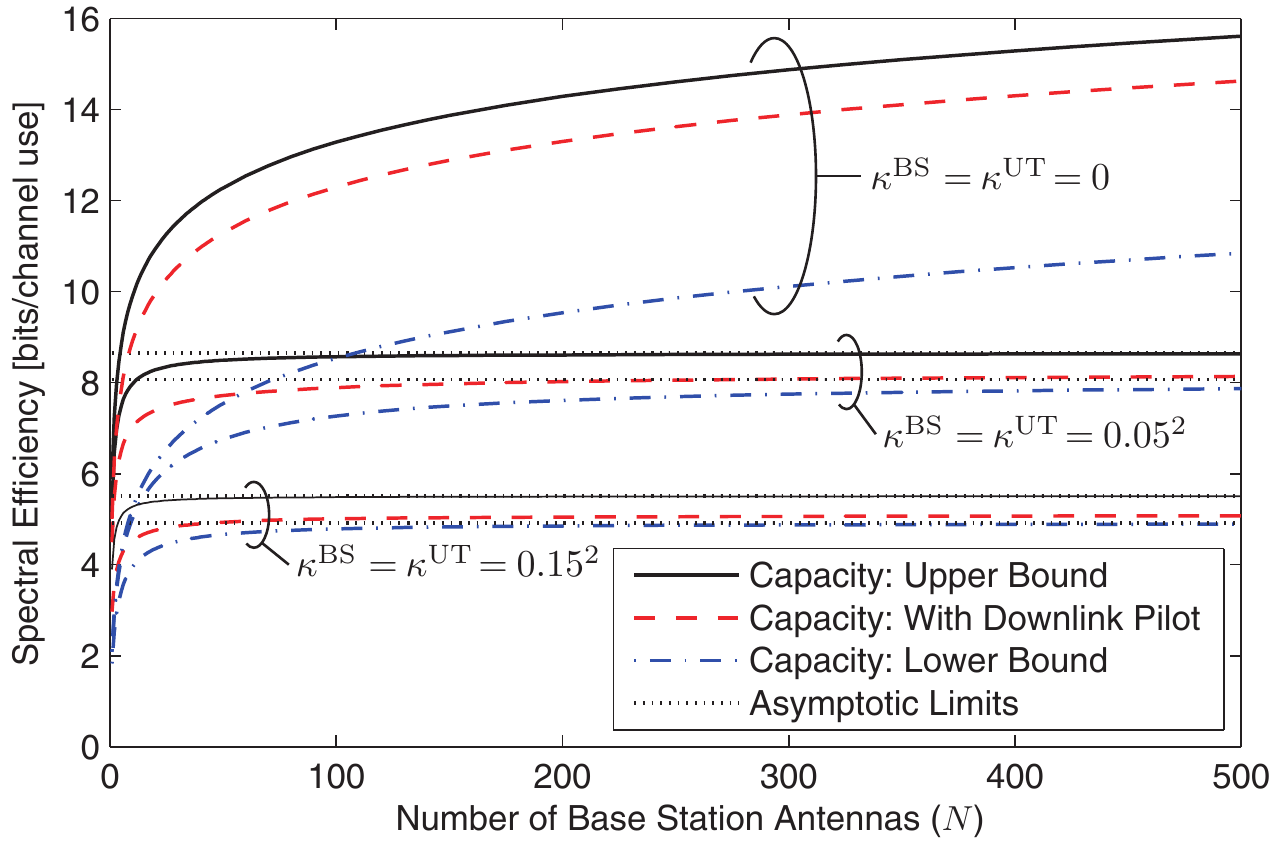} \vskip-3mm
\caption{Lower and upper bounds on the capacity. Hardware impairments have a fundamental impact on the asymptotic behavior as $N$ grows large.}\label{figure_capacity}
\end{center} \vskip-6mm
\end{figure}

Fig.~\ref{figure_capacity} considers a spatially uncorrelated scenario where $\vect{R}=\vect{S}=\vect{I}$.
While the capacity with ideal hardware grows without bound as $N \rightarrow \infty$, the lower and upper bounds converge to finite limits under hardware impairments. Recall that these bounds hold under any CSI conditions at the user terminal; for example, the dashed lines in Fig.~\ref{figure_capacity} show a lower bound for the practical case when a downlink pilot signal is transmitted for estimation of the effective channel $\vect{h}^T \frac{\hat{\vect{h}}}{\| \hat{\vect{h}} \|}$.

Observe that the asymptotic capacity limits in Fig.~\ref{figure_capacity} are characterized by the level of impairments, thus the hardware quality has a fundamental impact on the practical spectral efficiency. The majority of the multi-antenna gain is achieved at relatively low $N$; in particular, only minor improvements can be achieved by having more than $N=100$ antennas.

\begin{figure}
\begin{center}
\includegraphics[width=\columnwidth]{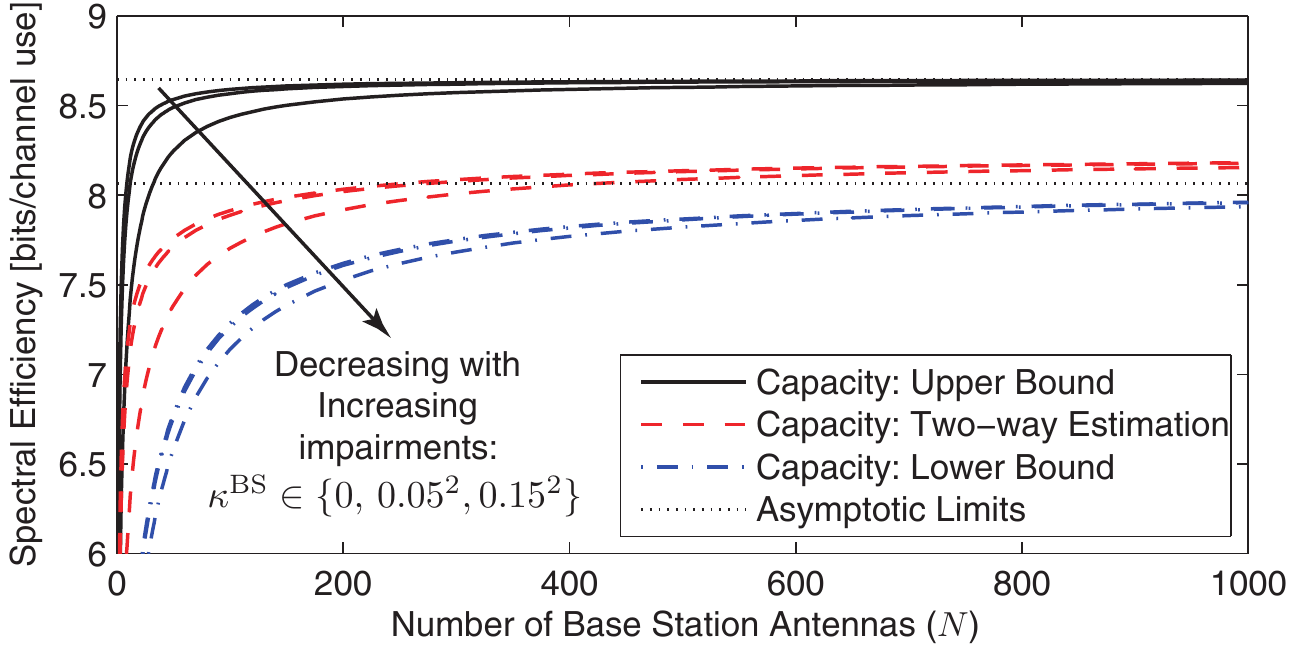} \vskip-2mm
\caption{Lower and upper bounds on the capacity for $\kappa^{\mathrm{UT}} = 0.05^2$. The impact of hardware impairments at the base station vanishes asymptotically.}\label{figure_capacity_diffimpairments}
\end{center} \vskip-4mm
\end{figure}

Next, Fig.~\ref{figure_capacity_diffimpairments} considers the same scenario but with a fixed level of impairments $\kappa^{\mathrm{UT}} = 0.05^2$ at the user terminal  and different values at the base station. As expected from the analysis, the lower and upper bounds at small $N$ decrease as $\kappa^{\mathrm{BS}}$ is increased, but the curves converge to the same value---thus showing that the impact of impairments at the base station vanishes as $N$ grows large.

\section{Energy Efficiency}
\label{sec:energy-efficiency}

This section analyzes the energy efficiency (in bits/Joule), which is defined as the ratio of capacity (in bits/channel use) to radiated power (in Joules/channel use) \cite{Ngo2013a}.

\setcounter{equation}{21}

\begin{definition}
The downlink energy efficiency is
\begin{equation} \label{eq:energy-efficiency}
{\tt EE} = \frac{{\tt C} }{p^{\mathrm{BS}} + \alpha_1 p^{\mathrm{BS}} + \alpha_2 p^{\mathrm{UT}}}
\end{equation}
where $\alpha_1,\alpha_2 \geq 0$ are some constants related to the overhead signaling in the downlink and uplink, respectively.
\end{definition}

Recall from Corollary \ref{cor:upper_bound} that the capacity has a finite ceiling as $N \rightarrow \infty$ or $p^{\mathrm{BS}} \rightarrow \infty$, thus high energy efficiency cannot be achieved by only increasing $N$ or $p^{\mathrm{BS}}$. However, we have the following important result, which has counterparts for ideal hardware \cite{Hoydis2013a,Ngo2013a} and the phase noise-impaired uplink \cite{Pitarokoilis2012a}.

\begin{corollary} \label{cor:energy-efficiency}
Suppose the downlink transmit power $p^{\mathrm{BS}}$ and uplink pilot power $p^{\mathrm{UT}}$ are scaled proportional to $1/N^{t_{\mathrm{BS}}}$ and $1/N^{t_{\mathrm{UT}}}$, respectively. If $t_{\mathrm{BS}} \geq 0, 0 < t_{\mathrm{UT}} < \frac{1}{2}$ and $t_{\mathrm{BS}}+t_{\mathrm{UT}}<1$, the lower bound in Corollary \ref{cor:lower_bound} gives asymptotically
\begin{equation} \label{eq:lower-bound-capacity-energy-eff}
\lim_{N \rightarrow \infty} {\tt C} \geq  \log_2 \left(1 + \frac{1 }{\kappa_r^{\mathrm{UT}} +\kappa_t^{\mathrm{UT}} + \kappa_r^{\mathrm{UT}} \kappa_t^{\mathrm{UT}} } \right).
\end{equation}
\end{corollary}
\begin{IEEEproof}
The dominated convergence theorem and $t_{\mathrm{UT}}>0$ gives $|\mathbb{E}\{ \cdot \}|^2 \rightarrow 1$ in the numerator and $\mathbb{E}\{ \cdot \} \rightarrow 1+\kappa_t^{\mathrm{UT}}$ in the denominator of \eqref{eq:lower-bound-capacity-energy-eff}. If $t_{\mathrm{BS}}+t_{\mathrm{UT}}<1$: $\frac{1}{N}\frac{\sigma_{\mathrm{UT}}^2}{p^{\mathrm{UT}} p^{\mathrm{BS}}} \rightarrow 0$.
\end{IEEEproof}

\begin{figure}
\begin{center}
\includegraphics[width=\columnwidth]{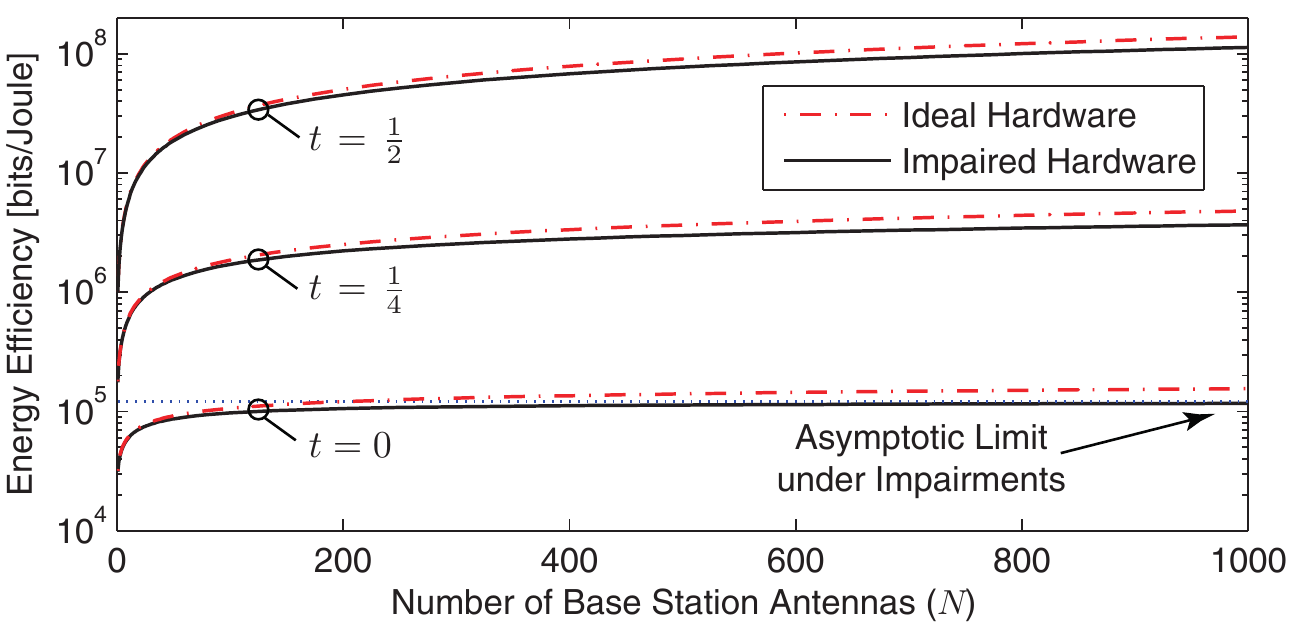} \vskip-2mm
\caption{Energy efficiency of ideal and impaired hardware when the transmit power and pilot powers are reduced with $N$ as $1/N^{t}$ for $t \in \{ 0, \frac{1}{4}, \frac{1}{2}\}$.}\label{figure_EE}
\end{center} \vskip-5mm
\end{figure}

Corollary \ref{cor:energy-efficiency} shows that one can reduce the transmit and pilot powers (e.g., roughly as $1/\sqrt{N}$) and still obtain a non-zero capacity. The asymptotic capacity is lower bounded by \eqref{eq:lower-bound-capacity-energy-eff}. As expected from previous results, the lower bound in \eqref{eq:lower-bound-capacity-energy-eff} only depends on the level of impairments at the user terminal. This implies a great robustness to base station impairments. The power scaling in Corollary \ref{cor:energy-efficiency} has the extraordinary consequence
\begin{equation} \label{eq:unbounded-EE}
\lim_{N \rightarrow \infty} {\tt EE} = \lim_{N \rightarrow \infty} \frac{{\tt C}}{  (1 + \alpha_1) \frac{p^{\mathrm{BS}}}{N^{t_{\mathrm{BS}}}} + \alpha_2 \frac{p^{\mathrm{UT}}}{N^{t_{\mathrm{UT}}}} } =
\infty.
\end{equation}
Observe that this unbounded asymptotic energy efficiency is achieved by using the LMMSE estimator in Theorem \ref{theorem:LMMSE-estimator} for uplink channel estimation and approximate MRT for downlink transmission. Large-scale MISO systems can thus obtain an immense energy efficiency even under hardware impairments.

\begin{remark}
Although the ratio of capacity to radiated power in \eqref{eq:energy-efficiency} can grow unboundedly as shown in \eqref{eq:unbounded-EE}, the overall energy efficiency (capacity divided by total power consumption) will not. Specifically, each antenna requires dedicated circuits with a non-zero power consumption $\varrho>0$. This contributes an additional term $N \varrho$ in the denominator of \eqref{eq:energy-efficiency} and implies that the overall energy efficiency is maximized at some finite $N$.
This refined efficiency notion was analyzed in \cite{Bjornson2013e} and it was observed that large-scale MIMO offers great energy savings also when the circuit power is taken into account.
\end{remark}

\subsection{Numerical Illustration}

The energy efficiency (using the lower bound in Corollary \ref{cor:lower_bound}) is illustrated in Fig.~\ref{figure_EE} when the transmit and pilot powers are $p^{\mathrm{BS}} = p^{\mathrm{UT}} = 30$ dBm at $N=1$ and then reduced by a factor $1/N^{t}$ for $t \in \{ 0, \frac{1}{4}, \frac{1}{2}\}$. The channel covariance matrices $\vect{R}$ are generated using the exponential model with correlation coefficient 0.7 \cite{Loyka2001a}, while the noise covariance  $\vect{S}$ is a scaled identity matrix such that $\tr(\vect{R}) / \tr(\vect{S}) = -10 $ dBm at a bandwidth of 15 kHz. The impairment parameters are set to $\kappa_t^{\mathrm{BS}} =\kappa_r^{\mathrm{BS}} = \kappa_t^{\mathrm{UT}} =\kappa_r^{\mathrm{UT}} = 0.05^2$.

Fig.~\ref{figure_EE} verifies that the energy efficiency grows unboundedly with $N$ for $t= \frac{1}{4}$ and $t=\frac{1}{2}$, while it converges to a finite value for $t=0$ (i.e., when the transmit/pilot powers are fixed). Note that $\alpha_1 = \alpha_2=0$ in the figure, but changing these parameters only result in a common vertical shift of all the curves.
Fig.~\ref{figure_EE} also shows the case of ideal hardware. Despite the fundamental differences when $N \rightarrow \infty$ or $p^{\mathrm{BS}},p^{\mathrm{UT}} \rightarrow \infty$ (i.e., there is no capacity ceiling or estimation error floor under ideal hardware), the difference in terms of energy efficiency is remarkably small between ideal and practical hardware.

\section{Conclusion}

This paper analyzed how transceiver hardware impairments impact the capacity and estimation accuracy of large-scale MISO systems. We proved analytically that the impairments of physical hardware create a finite capacity ceiling and non-zero estimation error floor---irrespective of the SNR and number of base station antennas $N$. This stands in contrast to the very optimistic asymptotic results previously reported for ideal hardware. Interestingly, only the hardware impairments at the user terminal limits the performance as $N$ grows large. Despite these discouraging results, we showed that large-scale MISO systems can achieve an arbitrarily high energy efficiency.

\bibliographystyle{IEEEtran}
\bibliography{IEEEabrv,refs_conf}

\end{document}